# Review of Equation of Motion in the Static Casimir Effect

Mohammad Mansouryar

## Abstract

*According to the experimentally observed theory of the static Casimir effect, two metal, uncharged, conductive and flat plates attract each other in vacuum. Herein, equations of motion of the plates which are influenced by Casimir potential are derived. Then some guess about experimental configurations that would provide possibility of observation of consequences of the derived results, are presented. As a result, frontiers of static and dynamic Casimir effects are somehow explored.*

## Introduction

The static Casimir effect is assumed to be originated of zero-point quantum vacuum energy fluctuations polarized by Dirichlet boundary conditions [1]. The following expression is given for the potential between two uncharged metallic conducting flat plates:

$$U_c(r) = \frac{-\mathbf{p}^2}{720} \frac{\hbar c A}{r^3} \quad , \quad F = -\frac{\partial}{\partial r} U \tag{1}$$

which shows the plates attract each other in zero Kelvin. This phenomenon has been measured with proper approximation in the laboratory [2,3,4]. In present paper, the equation of motion due to the potential $U(r)$ is derived and it is attempted to present some applications of it.

## Equation of Motion

If we consider motion of the plates in one dimension, for the classic and relativistic conditions we'll have respectively:

$$\ddot{r} - \frac{\mathbf{a}}{r^4} = 0 \quad , \quad r(0) = a \quad , \quad \dot{r}(0) = 0$$

$$\frac{\mathbf{a}}{r^4} = \ddot{r}\left[1 - \left(\frac{\dot{r}}{c}\right)^2\right]^{\frac{-1}{2}} + \frac{\dot{r}}{c^2}\left(\dot{r}\ddot{r}\left[1 - \left(\frac{\dot{r}}{c}\right)\right]^{\frac{-3}{2}}\right) \tag{2}$$

that $\mathbf{a} = \frac{-\mathbf{p}^2 \hbar c A}{240 m_0}$ and the plates supposed to be in an initial distance of "$a$" then start to attract each other. The solutions of the Eq. (2) have been approximated by Maple V:

$$r(t) = a + \frac{\mathbf{a}}{2a^4}t^2 - \frac{\mathbf{a}^2}{6a^9}t^4 + O(t^6) \quad , \quad v(t) = \frac{\mathbf{a}}{a^4}t - \frac{2\mathbf{a}^2}{3a^9}t^3 \quad Newtonian$$

$$r(t) = a + \frac{\mathbf{a}}{2a^4}t^2 - \frac{\mathbf{a}^2(4c^2a^3 + 3\mathbf{a})}{24c^2 a^{12}}t^4 - \frac{3\mathbf{a}^4}{40c^3 a^{16}}t^5 + O(t^6) \quad relativistic \tag{3}$$

now by substitution of numerical values of $\mathbf{a}$ and a, the motion can be analyzed. Let's assume $A = 1 \ cm^2$ and $m = 10^{-4} \ kg$ then $\mathbf{a} = .1300125998e\text{-}26$. In the case of the distance between the plates it is discussed that $10^{-6} m \leq a \leq 10^{-8} m$, but it is specified the Casimir force from the order of 10μm to plasma wavelength of the plates, is the dominant effect [5]. While $l_p = 2pc\left(\frac{m_e e_0}{Ne^2}\right)^{\frac{1}{2}} \approx \frac{4494690256}{N^{\frac{1}{2}}}$ then more particle density of the plates, less plasma wavelength and the plates would approach one another more, so that mass of the plates shouldn't be much because of more effectiveness of vacuum energy on them. For the metals, that is nearly equals to $0.4494 \times 10^{-10} m$. As described above, the metallic hydrogen plates with mass and area given above, their lattice spacing $d = \left(\frac{ALm_p}{M_p}\right)^{\frac{1}{3}}$ for thickness L of one layer of molecules is estimated 0.926 nm which is doubled for 8 layers. As reasoned in [6], whenever the zero-point field does positive work in a localized region of spacetime $W_\infty = \lim_{r_1 \to \infty}(W_{12} = -(U_2 - U_1))$, there would occur a falling in energy of electromagnetic vacuum zero-point field which results in a negative energy density in that region in respect to rest of the spacetime. As seen by (1), decreasing the distance between the plates, increasing the negative potential although for extraction of negative mass-energy or exotic matter [7,8] of Casimir system due to $m = \frac{U_\infty}{c^2} = -\frac{p^2}{720}\frac{\hbar A}{cr^3}$ in proper magnitudes, more arrangements are needed. Also, it is visionable in the nearest distance of holding the potential (1) due to direct dependence of the production rate of photon to frequency and frequency to velocity in the dynamical version of the Casimir effect [9], a photon emission in the static case can be expected. Although, that is merely a guess and final solution will be specified in the lab by satisfying the as ideal as possible conditions. As explanation, see formula (251) of Ref. [9] and consider $L(t) \propto \cos(wt) \Rightarrow \dot{L}(t) \propto -w\sin(wt)$ so although there are basic differences between the case of vibrating cavities –there is periodic motion and frequency if meaningful- and our under discussion case, but according to results presented in the last section of that paper, if the velocity of the boundary were of the order of "$c$", then a sufficient number of photons could be created from vacuum practically for any law of motion. Calculations based on Eq.(3) show the magnitude of plates' velocity has a strong dependence to the initial dependence they are located in, so that the less initial distance between them, their velocities would be increased when time passes much orders more than the conditions with more (initial) distance. As an example of a realistic situation, it can be assumed if the plates traverse the distance 10μm to 100nm, in (at most) 100ms, the velocity formulas gained of (3) are valid either, by accumulation of the process the photon production may be expected, see Fig. 1.

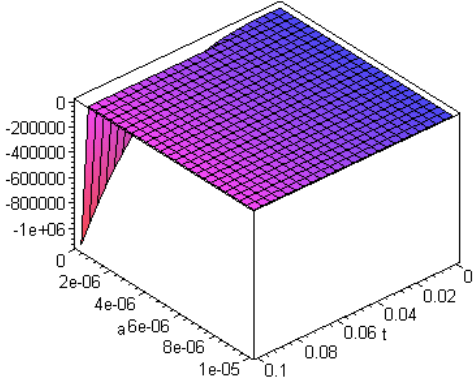

Fig 1: The velocity caused by Casimir potential for $t \in [0,0.1]$ and $a \in [10\,nm, 0.1\,mm]$.

### The Classic and Quantum approaches

As a matter of fact, the condition in Fig. 1 seem somehow ideal and in other possible intervals some problems in consideration of velocities caused by Casimir potential for the moving plates would raise. It means the theory results something that are not correct in the view of special relativity. This is actually originated of the fact that because of being much sensitive of the potential to changing of distance particularly in short distances, the classical review of this phenomenon is quite unacceptable. For stating the above comment quantitatively, exerting the Ehrenfest theorem in one dimension, we find $\Delta x = \langle x^2 \rangle - \langle x \rangle^2$ have to be small. In fact, by considering

$$\langle x^2 \rangle = \int_{x_1}^{x_2} u(x) x^2 u(x) dx \qquad \langle x \rangle^2 = \left[ \int_{x_1}^{x_2} u(x) x u(x) dx \right]^2 \tag{4}$$

it is desired to study the degree of being unacceptable of the classic approach for the Casimir potential. The $u(x)$ by inserting the potential in Shrödinger equation is derived that we have

$$U_C = V = \frac{-\mathbf{a}}{x^3} \Rightarrow \frac{d^2 u}{dx^2} + \frac{2m}{\hbar^2}(E - \mathbf{a}x^{-3}) u(x) = 0 \tag{5}$$

then the wave function of the problem must be calculated from the above equations. Besides, because the Eq.(5) cannot be solved by algebraic methods, by transforming $1/x = \mathbf{r}$ the equation changed to

$$\mathbf{r} \frac{d^2 u'}{d\mathbf{r}^2} - 2 \frac{du'}{d\mathbf{r}} + \mathbf{b}(E\mathbf{r}^5 - \mathbf{a}\mathbf{r}^8) u' = 0 \tag{6}$$

that its solution is

$$u'(\mathbf{r}) = C_1 \mathbf{r}^3 (1 + O(\mathbf{r}^6)) + C_2 (\ln(\mathbf{r})(O(\mathbf{r}^6)) + (12 + O(\mathbf{r}^6))) \tag{7}$$

and does not seem applicable but considering the solution about $x = 1$, by transforming the equation to ($\mathbf{b} = 2m/\hbar^2$, $x = t + 1$):

$$\frac{d^2 u}{dt^2} + \mathbf{b}\left[ E - \frac{\mathbf{a}}{(t+1)^3} \right] u(t) = 0 \tag{8}$$

we can gain the following approximation:

$$u(t) = u(0) + D(u)(0)\ t + \frac{\mathbf{b}}{2}(\mathbf{a} - E)u(0)\ t^2 + \left( \frac{-\mathbf{b}\mathbf{a}}{2} u(0) + \frac{\mathbf{b}\mathbf{a}}{6} D(u)(0) - \frac{\mathbf{b}E}{6} D(u)(0) \right) t^3 +$$

$$\left( \frac{\mathbf{b}^2 E^2}{24} u(0) - \frac{\mathbf{b}^2 \mathbf{a} E}{12} u(0) + \frac{\mathbf{b}^2 \mathbf{a}^2}{24} u(0) + \frac{\mathbf{b}\mathbf{a}}{2} u(0) - \frac{\mathbf{b}\mathbf{a}}{4} D(u)(0) \right) t^4 +$$

$$\left( \frac{-\mathbf{b}\mathbf{a}}{2} u(0) + \frac{3\mathbf{b}\mathbf{a}}{10} D(u)(0) - \frac{\mathbf{b}^2 \mathbf{a}^2}{10} u(0) + \frac{\mathbf{b}^2 \mathbf{a}^2}{120} D(u)(0) - \frac{\mathbf{a}\mathbf{b}^2 E}{60} D(u)(0) + \frac{\mathbf{b}^2 \mathbf{a} E}{10} u(0) + \frac{\mathbf{b}^2 E^2}{120} D(u)(0) \right) t^5 + O(t^6)$$

$$\tag{9}$$

## Conclusion

In fact, by realistic assumptions about involved parameters in the Eq.(9), inserting in Eq.(4), adjusting $x_1, x_2$, we can gain relations between manually modifiable variables and precision of the approach presented here for dealing with the Casimir effect.